\documentclass[namedreferences]{solarphysics}

\usepackage[hyperref,optionalrh,natbib]{spr-sola-addons} 
\usepackage{graphicx}        
\usepackage{color}           

\ifx \doiurl    \undefined \def \doiurl#1{\href{http://dx.doi.org/#1}{\textsf{DOI}}}\fi
\ifx \adsurl    \undefined \def \adsurl#1{\href{http://adsabs.harvard.edu/abs/#1}{\textsf{ADS}}}\fi
\ifx \arxivurl  \undefined \def \arxivurl#1{\href{http://arxiv.org/abs/#1}{\textsf{arXiv}}}\fi

\newcommand{\apj}{    {\it Astrophys. J.}}
\newcommand{\apjl}{   {\it Astrophys. J. Lett.}}

\newcommand{\pasj}{   {\it Publ. Astron. Soc. Japan}}

\newcommand{\solphys}{{\it Solar Phys.}}

\begin{document}

\begin{article}

\begin{opening}

\title{Estimating Flare-related Photospheric Lorentz Force Vector Changes within Active Regions}

\author{G.J.D.~\surname{Petrie}      
       }
\runningauthor{Petrie}
\runningtitle{Estimating Flare-related Lorentz Force Changes}

  \institute{National Solar Observatory, Tucson, AZ 85719, USA.
                     email: \url{gpetrie@nso.edu}\\
             }

\begin{abstract}
It is shown that expressions for the global Lorentz force associated with a flaring active region derived by Fisher {\it et al.} (2012) can be used to estimate the Lorentz force changes for strong fields in large structures over photospheric subdomains within active regions. Gary's~(2001) model for the stratified solar atmosphere is used to demonstrate that in large-scale structures with typical horizontal magnetic length scale $\gg 300$~km and with strong magnetic fields ($\ge 1$~kG at the $\tau =1$ opacity layer at 5000~\AA), the Lorentz force acting on the photosphere may be approximated by a surface integral based on photospheric boundary data alone. These conditions cover many of the sunspot fields and major neutral lines that have been studied using Fisher {\it et al.}'s (2012) expressions over the past few years. The method gives a reasonable estimate of flare-related Lorentz force changes based on photospheric magnetogram observations provided that the Lorentz force changes associated with the flare have a lasting effect on the observed fields, and are not immediately erased by post-flare equilibration processes.

\end{abstract}
\end{opening}

\section{Introduction}
\label{sect:intro}

In recent years it has repeatedly been shown that the solar photosphere responds to the abrupt coronal reconfiguration of most major flares, often with abrupt, coherent, permanent, and widely distributed patterns of photospheric magnetic field change. Flare-related photospheric magnetic field changes were sought for many years ({\it e.g.}, Severny,~1964; Zirin and Tanaka,~1981), before abrupt, permanent photospheric field changes were successfully observationally linked to flares; see the discussion in Sudol and Harvey~(2005). Since the launch of NASA's {\it Solar Dynamics Observatory} (SDO) satellite (Pesnell {\it et al.},~2012), the {\it Helioseismic and Magnetic Imager} (HMI) instrument (Schou {\it et al.},~2011) has provided full-disk vector magnetograms with $0.\!\!^{\prime\prime}5$ pixels every 12 min, enabling the study of flare-related photospheric vector magnetic field changes in more detail than was possible previously.

The history of flare-related photospheric magnetic field changes has not been straightforward. Following the groundbreaking work of Wang {\it et al.}~(1992, 1994), reporting abrupt, permanent field changes in flaring active regions, a number of later studies produced inconclusive results; see the discussion in Wang~(2006). Then, the introduction of high-cadence longitudinal (line-of-sight) photospheric magnetic field measurements, from NASA's {\it Michelson Doppler Imager} (MDI) on the {\it Solar and Heliospheric Observatory} (SoHO) satellite and the National Solar Observatory’s {\it Global Oscillations Network Group} (GONG) ground-based network, resulted in mounting evidence of flare-related magnetic changes in the photosphere. Based on one-min MDI longitudinal data, Kosovichev and Zharkova~(1999, 2001) reported sudden and permanent decreases in magnetic flux near X-class flares near magnetic neutral lines, and Zharkova {\it et al.}~(2005) found a permanent and significant increase in magnetic flux near the neutral line of a region during another X-class flare observed near the limb. Using one-min GONG longitudinal magnetograms, Sudol and Harvey~(2005) characterized the spatial distribution, strength, and rate of change of permanent field changes associated with 15 X-class flares. By carefully co-registering the images they succeeded in tracing the field changes pixel-by-pixel and were able to show the spatial structure of the changes. They concluded that the majority of field changes occurred close to or within sunspots. Wang~(2006) found an unshearing movement parallel to the neutral lines in flare-related longitudinal magnetic field changes in the MDI data for all five $\delta$-spot flares that he studied, implying an overall release of shear, but that the two polarities converged toward the neutral line during some events and diverged during others. Petrie and Sudol~(2010) analyzed 1-min GONG longitudinal magnetograms covering 77 flares of GOES class at least M5 and, exploring the relationship between increasing/decreasing longitudinal fields and azimuth and tilt angles at various positions on the disk, found that the overall distributions of longitudinal increases and decreases at different parts of the disk was found to be consistent with the coronal implosion interpretation (Hudson,~2000; Hudson, Fisher, and Welsch,~2008; Fisher {\it et al.},~2012) where, after a coronal magnetic eruption, the remaining coronal field loop structure contracts downward resulting in a more horizontal photospheric field.

Wang and Liu~(2010) studied 11 X-class flares using vector magnetograms from the vector magnetograph of Big Bear Solar Observatory and other vector magnetographs, and found in each case an increase of transverse field at the main polarity-inversion line. The launch of SDO and the release of 12-min HMI photospheric vector magnetograms prompted many authors to study flare-related photospheric vector field changes. Wang {\it et al.}~(2012), Gosain~(2012), Sun {\it et al.}~(2012) and Petrie~(2012, 2013) analyzed the 12-min HMI vector data for the 15 February 2011 X2.2 flare using different methods, and again found an increase of transverse field at the polarity-inversion line, as did Liu {\it et al.}~(2012) for the 13 February 2011 M6.6 flare. These observations support Hudson, Fisher, and Welsch’s~(2008) loop-collapse scenario.

Hudson, Fisher, and Welsch~(2008) and Fisher {\it et al.}~(2012) developed a method for estimating the Lorentz force change acting on the photosphere from the atmosphere above using the changes in the photospheric vector field observed during (or shortly before and after) the flare. The Lorentz force is estimated by integrating components of the Maxwell stress tensor across the boundary of a large domain containing the active region (see Figure~1 of Fisher {\it et al.},~2012). The bottom boundary of the domain is identified with the photospheric layer from which the observations derive, and the field values on this boundary are represented by the photospheric magnetic field observations. Because the field values at the lateral and top boundaries of the domain are not measured, Fisher {\it et al.}~(2012) advocated applying their method only to domains where these boundaries are far enough away from the region that the fields crossing them are too weak to contribute significantly to the integral, allowing one to ignore them. Strict adherence to this approach eliminates the possibility of studying the diverse responses of distinct components of an active region, such as sunspots and neutral-line arcades, whose behavior may provide important clues regarding the physics of the flare. Because of this, some authors have proceeded to apply Fisher {\it et al.}'s~(2012) equations to subdomains within active regions ({\it e.g.}, Wang {\it et al.},~2012; Petrie,~2012, 2013), sometimes providing brief arguments in favor of doing so (Alvarado-G\'omez {\it et al.},~2012; Petrie,~2013). In particular, it has been argued that the force-free nature of the coronal magnetic field allows one to neglect most of the contribution from the lateral and top boundaries of such a domain. In this paper we will revisit this problem using a model for the stratified solar atmospheric field.

When the 12-min HMI vector data became available it became possible to apply Fisher {\it et al.}'s~(2012) surface integrals for the photospheric Lorentz force, and it is noteworthy that these calculations have produced strikingly consistent results. Wang {\it et al.}~(2012) found that the vertical Lorentz force changes associated with 18 major flares were generally directed downward near neutral lines and were correlated with peak soft X-ray flux. Petrie~(2012) studied abrupt changes in both the photospheric magnetic and Lorentz force vector associated with six major neutral-line flares using HMI vector data. During all six flares the neutral-line field vectors became stronger and more horizontal, in each case almost entirely due to strengthening of the horizontal field components parallel to the neutral line. In all cases the neutral-line pre-flare fields were less tilted than the reference potential fields, and collapsed abruptly and permanently closer to potential-field tilt angles, implying that the relaxation of magnetic stress played a leading role in creating the magnetic changes. Indeed, the vertical Lorentz force had a large, abrupt, permanent downward change at the main neutral line during each of the flares, consistent with loop collapse. The horizontal Lorentz force changes acted mostly parallel to the neutral line in opposite directions on each side, a signature of the fields contracting during this loop collapse, pulling the two sides of the neutral line toward each other. The greater effect of the flares on field tilt than on shear may be explained by photospheric line-tying, since shear cannot be removed during loop collapse without the foot-points moving across the photosphere. Petrie~(2013) found that, in the case of the 15 February 2011 X2.2 flare, the oppositely-directed horizontal Lorentz force changes acting on each side of neutral line were accompanied by abrupt torsional un-twisting forces in the two sunspots at each end of the neutral line.

While these studies exploited Fisher {\it et al.}'s (2012) equations to offer more physical insight into the flare-related magnetic changes than the magnetic vector information alone, it remained true that Fisher {\it et al.}~(2012) did not advocate applying their method to fields within active regions and that a detailed defense was lacking. The purpose of this paper is to investigate whether and under what circumstances Lorentz force estimates within active regions can be reliably produced using this method. In this paper we revisit Fisher {\it et al.}'s calculation in the context of Gary's~(2001) well-known model for the plasma $\beta$ (the ratio between the plasma and magnetic pressures) of an active region. This allows us to study the influence of solar atmospheric stratification on Lorentz-force distribution between the photosphere and the corona.

The paper is organized as follows. In Section~\ref{sect:fisher} we review Fisher {\it et al.}'s~(2012) method. In Section~\ref{sect:subdomain} we derive expressions representing the contributions to the Lorentz force integral from the different boundary surfaces of a subdomain within an active region, and arrive at inequalities describing conditions where photospheric Lorentz forces within active regions can be derived from photospheric measurements alone. In Section~\ref{sect:gary} we introduce Gary's~(2001) plasma $\beta$ model and in Section~\ref{sect:lfintegral} we estimate force contributions from different atmospheric layers using this model. In Section~\ref{sect:conclusion} we arrive at conclusions regarding the application of Fisher {\it et al.}'s method to subdomains within active regions.

\section{Fisher {\it et al.}'s~(2012) Method for Estimating Photospheric Lorentz Force Changes Associated with a Flaring Active Region}
\label{sect:fisher}

We briefly summarize the method of Fisher {\it et al.}~(2012) to estimate the total Lorentz force vector change acting on the photosphere and subphotospheric volume below an active region as a result of a flare. The Lorentz force per unit volume ${\bf f}_{\rm L}$ can be written as,

\begin{equation}
{\bf f}_{\rm L} = {\bf\nabla}\cdot{\bf T} = {\partial T_{ij}\over{\partial x_j }} ,
\label{eq:vollorentz}
\end{equation}

\noindent where the Maxwell stress tensor $(T_{ij})$ in local Cartesian coordinates in terms of the spherical field components $(B_r, B_{\theta}, B_{\phi} )$ is,

\begin{equation}
{\bf T} = \frac{1}{8\pi}\left[
\begin{array}{ccc}
B_r^2-B_{\theta}^2-B_{\phi}^2 & 2B_rB_{\theta} & 2B_rB_{\phi} \\
2B_rB_{\theta} & B_{\theta}^2-B_r^2-B_{\phi}^2 & 2B_{\theta}B_{\phi} \\
2B_rB_{\phi} & 2B_{\theta}B_{\phi} & B_{\phi}^2-B_r^2-B_{\theta}^2
\end{array}\right] .
\label{eq:maxwelltensor}
\end{equation}

Fisher {\it et al.}~(2012) evaluated the total Lorentz force over an atmospheric volume surrounding an isolated flaring active region by integrating Equation~(\ref{eq:vollorentz}) over this volume, whose lower boundary is identified with the photosphere, with upper boundary far above the photosphere, and side boundaries connecting these surfaces to form a closed volume $V$ as shown in Figure~1 of Fisher {\it et al.}~(2012). Evaluating the volume integral of Equation~(\ref{eq:vollorentz}) using Gauss's divergence theorem then gives (Fisher {\it et al.},~2012),

\begin{equation}
{\bf F}_{\rm L} = \int_V {\bf\nabla}\cdot{\bf T}\ \mathrm{d}V = \int_{A_\mathrm{tot}} {\bf T}\cdot{\bf\hat{n}}\ \mathrm{d}A,
\label{eq:gaussdiv}
\end{equation}

\noindent where the area integral is evaluated over all surfaces of the volume, denoted by $A_\mathrm{tot}$, with unit normal vector ${\bf\hat{n}}$. As Fisher {\it et al.}~(2012) argue, if the upper boundary of the volume is far enough above the photosphere and the side boundaries are distant enough from the active region that the field integrals over these surfaces are negligible, then the surface integral of Equation~(\ref{eq:gaussdiv}) reduces to an integral over the photospheric lower boundary $A_\mathrm{ph}$ only. In this case, for the force acting on the volume below the photosphere, ${\bf\hat{n}} = {\bf\hat{r}}$ and ${\bf B}\cdot{\bf\hat{n}} = B_r$ and,

\begin{equation}
F_r = {1 \over 8 \pi} \int_{A_\mathrm{ph}} ( B_r^2 - B_h^2 )\ \mathrm{d}A,
\label{eq:fr}
\end{equation}
and
\begin{equation}
{\bf F}_h = {1 \over 4 \pi} \int_{A_\mathrm{ph}} ( B_r {\bf B}_h )\ \mathrm{d}A,
\label{eq:fh}
\end{equation}

\noindent where we have reversed the signs of these expressions compared to Fisher {\it et al.}'s~(2012) Equations~(5) and (6) because we are considering the forces imposed on the photosphere from above instead of the equal and opposite forces on the atmosphere from below (Fisher {\it et al.},~2012).

If one sets the total Lorentz-force to zero, the above surface integrals yield well-known expressions for the Lorentz force on the photosphere (Molodensky,~1974) that have been used to derive conditions for the force-freeness of measured photospheric magnetic vector fields (Low,~1985). These expressions have also been used to compare the forces of measured fields in the low solar satmosphere using photospheric and chromospheric vector magnetograms (Metcalf {\it et al.}, 1995). Now, assuming that the photospheric vector field is observed over a photospheric area $A_\mathrm{ph}$ at two times, $t=0$ before the flare-related field changes begin, and $t=\delta t$ after the main field changes have occurred, the corresponding changes in the Lorentz force vector components between these two times are given by Equations~(17) and (18) of Fisher {\it et al.}~(2012), again with a change of sign:

\begin{equation}
\delta F_r = {1 \over 8 \pi} \int_{A_\mathrm{ph}} ( \delta B_r^2 - \delta B_h^2 )\ \mathrm{d}A,
\label{eq:deltafr}
\end{equation}
and
\begin{equation}
\delta {\bf F}_h = {1 \over 4 \pi} \int_{A_\mathrm{ph}} \delta ( B_r {\bf B}_h )\ \mathrm{d}A,
\label{eq:deltafh}
\end{equation}
where at a fixed location in the photosphere
\begin{eqnarray}
\delta B_h^2 & = & B_h^2(\delta t) -B_h^2(0)\ ,\\
\delta B_r^2 & = & B_r^2(\delta t) -B_r^2(0)\ ,\\
\delta ( B_r {\bf B}_h ) & = & B_r(\delta t){\bf B}_h(\delta t) - B_r(0) {\bf B}_h(0)\ .
\end{eqnarray}


\section{Estimating Lorentz Force Changes within a Subdomain of an Active Region}
\label{sect:subdomain}

Let us revisit the Lorentz force integral Equation~(\ref{eq:gaussdiv}), this time considering the contributions not only from $A_{\rm ph}$ but from all of $A_{\rm tot}$, with the components of ${\bf T}$ given in Equation~(\ref{eq:maxwelltensor}). In Equation~(\ref{eq:gaussdiv}), ${A_\mathrm{tot}}$ comprises the photospheric boundary ${A_\mathrm{phot}}$ plus the lateral boundaries ${A_\mathrm{lat}}$ and the top boundary ${A_\mathrm{top}}$.

We now consider the case of a regular sub-domain within the active region with lower boundary at the photosphere. According to Equation~(\ref{eq:gaussdiv}) the contribution to the Lorentz force vector components from a rectangular photospheric sub-domain of horizontal dimensions $l_x\times l_y$ has typical size,

\begin{equation}
F_{\rm phot} = \frac{1}{8\pi} \int_{y=-l_y}^{l_y} \int_{x=-l_x}^{l_x} B^2(x,y,h_0) {\rm d} x{\rm d} y \approx \frac{B_0^2}{8\pi} 4l_x l_y=\frac{B_0^2}{2\pi} l_x l_y,\label{eq:photbdy}
\end{equation}

\noindent where $z=h_0$ is the height of the photospheric layer from where the observations derive, and $B_0$ is a typical field strength at that height. Here we have represented the Maxwell stress tensor components in Equation~(\ref{eq:maxwelltensor}) by the magnetic pressure alone for simplicity. This approximation neglects the effects of magnetic tension but is likely to give a useful description of the force distribution because both the magnetic pressure and tension forces are proportional to the square of the field strength.


Unlike the photospheric boundary, the lateral boundaries extend to great heights. On the other hand, recall that for a force-free field the components of the Maxwell stress tensor Equation~(\ref{eq:maxwelltensor}) integrate to zero (Molodensky,~1974; Low,~1985). Above a certain height $z=L$ in the atmosphere we expect the magnetic field to be close enough to a force-free state, or to be weak enough, that its contribution to Equation~(\ref{eq:gaussdiv}) is negligible compared to the contribution from the fields below $z=L$. Here we will assume that this is true, and we will test this assumption in Section~\ref{sect:lfintegral} using models for the stratification of the field.

For simplicity we confine our attention to large, coherent magnetic structures where the field components $B_i$ vary over a length scale $L_B$ such that

\begin{equation}
\frac{\partial B_i}{\partial x} \approx \frac{B_i}{L_B},\ \ \ \frac{\partial B_i}{\partial y} \approx \frac{B_i}{L_B}. \label{eq:horizderivs}
\end{equation}

\noindent In this case the difference between the Lorentz forces at the two lateral boundaries at $x=-l_x$ and $x=l_x$ at height $z$ has typical size

\begin{equation}
\frac{\Delta_x (B^2)}{8\pi} \approx \frac{2 B_{\rm ave} (z)}{8\pi} . \frac{2l_x B_{\rm ave} (z)}{L_B} = \frac{l_x B_{\rm ave} (z)^2}{2\pi L_B} ,
\label{eq:deltaB2}
\end{equation}

\noindent where $B_{\rm ave}(z)$ is a typical field strength at any height $z$. Then, defining $L$ to be the height over which the field has significant Lorentz forces, the typical size of contribution to the surface integral Equation~(\ref{eq:gaussdiv}) from two opposite lateral boundaries located at $x=\pm l_x$ is,

\begin{eqnarray}
F_{\rm lat} & = & \frac{1}{8\pi} \int_{z=h_0}^{L} \int_{y=-l_y}^{l_y} \frac{\Delta_x (B^2)}{8\pi} {\rm d} y {\rm d} z \nonumber\\
& \approx & \frac{l_x l_y}{\pi L_B} \int_{z=h_0}^{L} B_{\rm ave}(z)^2 {\rm d} z \label{eq:latbdy}.
\end{eqnarray}

\noindent The expression for the typical size of the contribution to the surface integral Equation~(\ref{eq:gaussdiv}) from the pair of lateral boundaries at $y=-l_y$ and $y=l_y$, based on the difference ${\Delta_y (B^2)}/(8\pi )$ between the Lorentz forces at $y=-l_y$ and $y=l_y$ at height $z$, is identical.  The contributions from the two pairs of lateral boundaries are the same because, from Equations~(\ref{eq:horizderivs}), the horizontal magnetic length scale is (approximmately) the same, $L_B$, in both horizontal directions, and because, from Equation~(\ref{eq:latbdy}), $F_{\rm lat}$ is proportional to both $l_x$ and $l_y$.

Comparing Equations~(\ref{eq:photbdy}) and (\ref{eq:latbdy}), the lateral surface contributions are small enough to be neglected if $F_{\rm lat} \ll  F_{\rm phot}$, {\it i.e.}, if, 

\begin{equation}
\frac{I_z(L)}{L_B} \ll  \frac{B_0^2}{2},\label{eq:latcomp}
\end{equation}

\noindent where
 
\begin{equation}
I_z(L) = \int_{z=h_0}^{L} B_{\rm ave}(z)^2 {\rm d} z.\label{eq:izdef}
\end{equation}

The comparison in Equation~(\ref{eq:latcomp}) is independent of the horizontal dimensions $l_x$ and $l_y$ of the domain of integration. So long as the magnetic structure is large and coherent enough that the condition Equation~(\ref{eq:latcomp}) is met, the Lorentz force can be approximated neglecting the lateral boundary contributions.

It remains to consider the contribution to Equation~(\ref{eq:gaussdiv}) from the top boundary at $z=L$, which has typical size

\begin{equation}
F_{\rm top} = \frac{1}{8\pi} \int_{y=-l_y}^{l_y} \int_{x=-l_x}^{l_x} B^2(x,y,L) {\rm d} x{\rm d} y \approx \frac{B_{\rm top}^2}{8\pi}.4l_x l_y = \frac{B_{\rm top}^2}{2\pi} l_x l_y ,\label{eq:topbdy}
\end{equation}

\noindent where $B_{\rm top}$ is a typical field strength at the height $z=L$. Therefore, the contribution from the top boundary can be neglected if $F_{\rm top} \ll  F_{\rm phot}$, {\it i.e.}, if

\begin{equation}
B_{\rm top}^2 \ll  B_{\rm phot}^2.
\label{eq:topcomp}
\end{equation}

Equations~(\ref{eq:latcomp}) and (\ref{eq:topcomp}) describe conditions where the lateral and top boundary contributions to the Lorentz force integral Equation~(\ref{eq:gaussdiv}) are much smaller than the photospheric contributions, and can be neglected, so that the whole-surface integrals can be approximated by the lower boundary integrals alone. Under these conditions, a reasonable Lorentz force estimate can be performed using photospheric measurements alone. Equations~(\ref{eq:latcomp}) and (\ref{eq:topcomp}) represent criteria that we will use to test various model fields in Section~\ref{sect:lfintegral} of the paper.

Clearly, not all fields pass this test. For example, with any force-free field the three components of Equation~(\ref{eq:gaussdiv}) must be zero, and these integrals therefore cannot possibly have dominant lower boundary contributions; $F_{\rm phot}$ cannot be much larger than both $F_{\rm lat}$ and $F_{\rm top}$. In these cases, therefore, the lateral and top boundary integrals cannot be neglected, and the lower boundary integral cannot accurately represent the whole boundary integral. A good Lorentz force estimate therefore cannot be derived from photospheric measurements alone for force-free fields. To satisfy Equations~(\ref{eq:latcomp}) and (\ref{eq:topcomp}), the flux distribution must be very different from that of a force-free field: the field strength must fall off sufficiently sharply with height that the photospheric contribution to the force integral Equation~(\ref{eq:gaussdiv}) is dominant. With flux conservation this means that the field must contain significant return flux (escape through the bottom boundary) and/or flux tube expansion (escape through the lateral boundaries). But the details of how the flux escapes from the domain to achieve this stratification are not directly relevant to Equations~(\ref{eq:latcomp}) and (\ref{eq:topcomp}). The condition imposed by Fisher {\it et al.}~(2012), that the field integrals over the top and lateral boundaries must be negligible, is a sufficient but not a necessary condition for Equations~(\ref{eq:latcomp}) and (\ref{eq:topcomp}) to be fulfilled. Equations~(\ref{eq:latcomp}) and (\ref{eq:topcomp}) can be fulfilled within an active region if the field strength decreases with height fast enough and the lateral boundary pressure differences are not too large. In Section~\ref{sect:lfintegral} we will use models to determine whether fields can be sufficiently steeply stratified under solar conditions that both of Equations~(\ref{eq:latcomp}) and (\ref{eq:topcomp}) are satisfied for reasonable horizontal length scales $L_B$.

Before we introduce models, we can note that the solar atmospheric field strength does indeed fall off sharply according to various observations summarized by Gary~(2001), and this is what is needed to satisfy Equations~(\ref{eq:latcomp}) and (\ref{eq:topcomp}). Moreover, Metcalf {\it et al.}~(1995) estimated that the atmosphere is approximately force-free ({\it i.e.}, much less forced than the photosphere) about 400~km above the photosphere. This result suggests that a box of integration with height $L=400$~km would capture all significant Lorentz forces. Crudely estimating $I_z(L) \le B_0^2 L$ from Equation~(\ref{eq:izdef}), Equation~(\ref{eq:latcomp}) then says that for magnetic structures of characteristic horizontal length scale $L_B \gg  2L = 800$~km, the contribution $F_{\rm lat}$ to Equation~(\ref{eq:gaussdiv}) should be negligible. This crude estimate leads us to expect that many photospheric Lorentz forces may be estimated using photospheric field measurements alone, but it does not adequately take into account the effects of solar atmospheric stratification. Not only does it ignore the effect of the photospheric fields generally being much stronger than most of the fields crossing the lateral boundary, but also the small but finite contributions to the Lorentz forces from fields in and above the chromosphere. 

To explore the influence of solar atmosphere stratification on the validity of the result, it is necessary to rely on a model for the stratification of the solar atmospheric field. Current extrapolation solutions are too weakly stratified to reproduce the stratification of the solar atmosphere from the photosphere to the corona. This applies not only to force-free fields that we discussed above, but also to non-force-free fields. Non-force-free extrapolations are rare, but Hu {\it et al.}~(2010) presented such a model based on an optimized linear combination of a potential field and two linear force-free fields. This model may represent the Lorentz forces reasonably well at the height of the observations in the photosphere, but not the great differences between the measured photospheric fields and the nearly force-free fields only a few hundred km above. To determine whether Fisher {\it et al.}'s~(2012) equations can give useful estimates of major photospheric Lorentz force changes within active regions using photospheric data, it is essential to represent the solar stratification accurately. We need another approach. In this paper we apply instead the simple and observationally constrained model of solar atmospheric stratification presented by Gary~(2001).

\section{Gary's~(2001) Model for the Magnetic and Plasma Stratification above an Active Region}
\label{sect:gary}

\begin{figure} 
\begin{center}
\resizebox{\textwidth}{!}{\includegraphics*{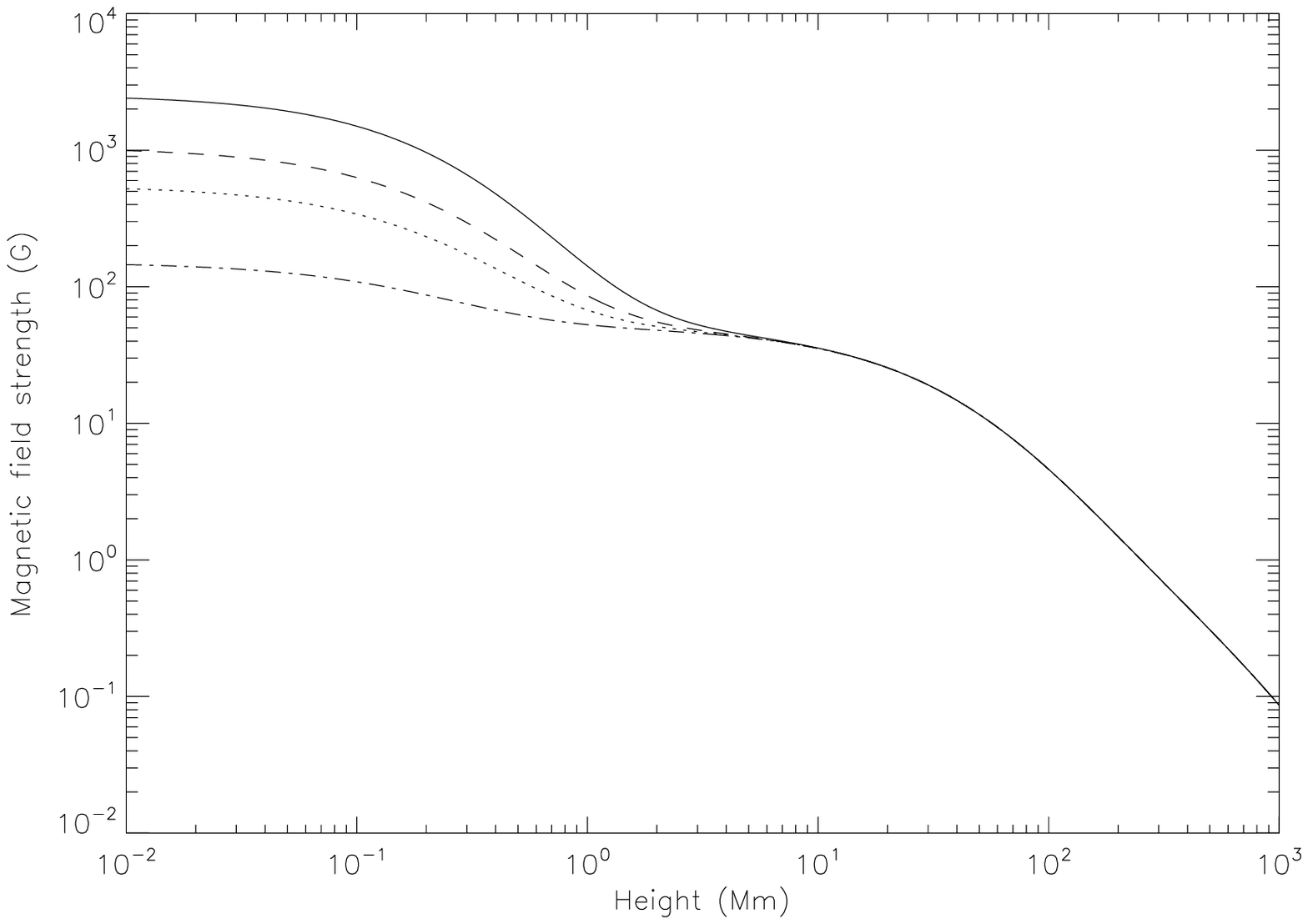}}
\resizebox{\textwidth}{!}{\includegraphics*{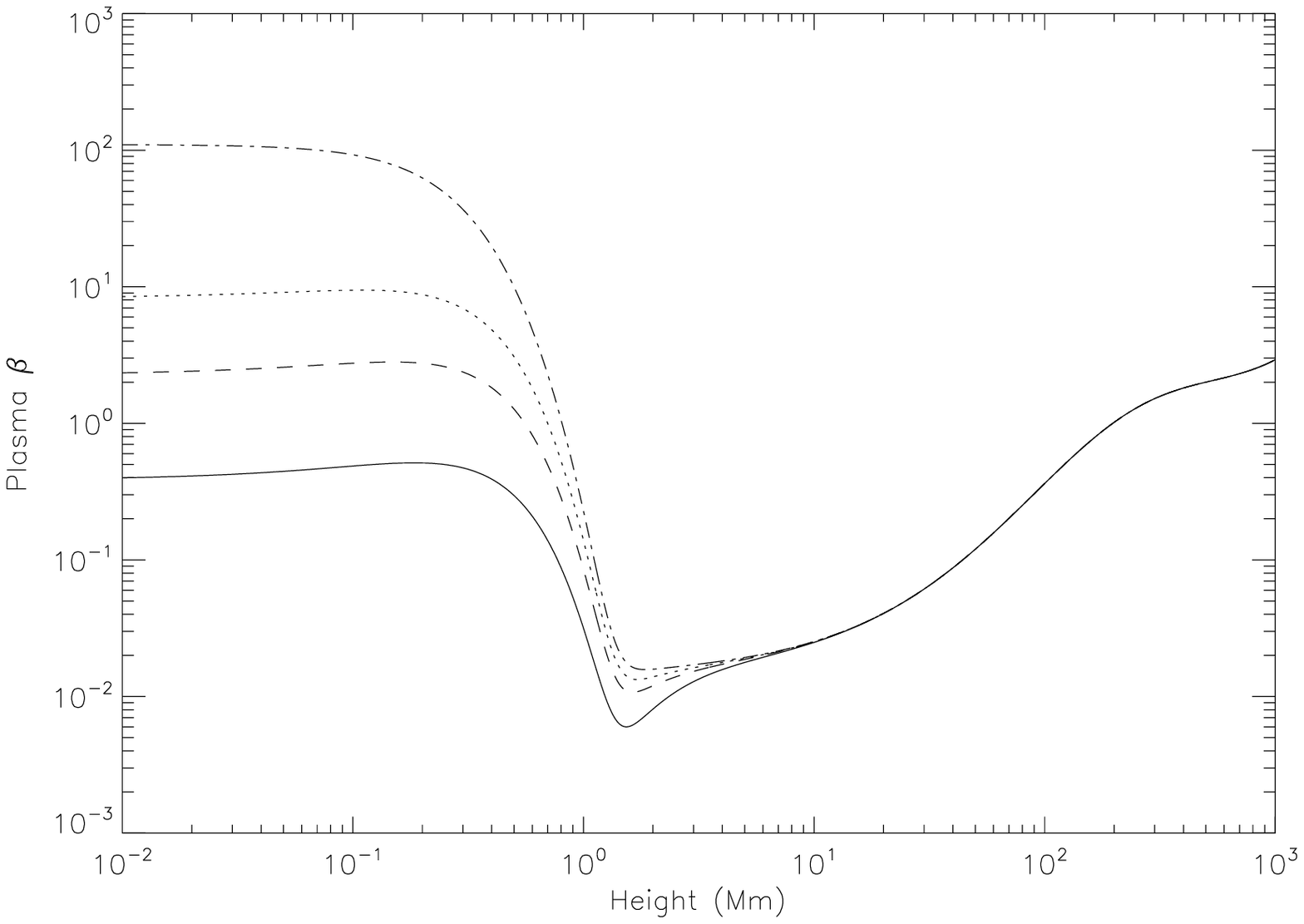}}
\end{center}
\caption{The magnetic field strength (top) and plasma beta (bottom) for four magnetic field models, plotted against height in the solar atmosphere. The solid and dot-dashed lines represent the models for sunspot and plage fields with photospheric field strengths of 2500~gauss (G) and 100~G presented by Gary~(2001). The dashed and dotted lines represent intermediate fields of photospheric field strength 1000~G and 500~G.}
\label{fig:gary}
\end{figure}

The solar corona has a complex structure with the plasma pressure and magnetic pressure dominant in different domains. The plasma pressure tends to be very dominant in the solar interior, though this may not be the case in some intense sunspot fields near the surface. The magnetic pressure tends to dominate between the chromosphere and mid-corona, though some atmospheric structures such as weak-field prominences seem likely to have significant plasma forces. In the open-field domain of the high corona the thermal and kinetic pressures of the expanding solar wind dominate the magnetic pressure, forcing the field to open at heights around a solar radius. The dimensionless parameter that characterizes this competition between the magnetic pressure $B^2/8\pi$ and plasma pressure $p$ is the plasma $\beta$,

\begin{equation}
\beta =\frac{8\pi p}{B^2}.
\label{eq:beta}
\end{equation}

\noindent Generally, $\beta > 1$ in the photosphere and $\beta < 1$ between the chromosphere and mid-corona, and becomes $> 1$ again around a solar radius above the photosphere (Gary,~2001).

Estimates of the relative contributions to the force integral Equation~(\ref{eq:gaussdiv}) from photospheric compared to non-photospheric heights must account for both the decrease in magnetic field strength as a function of height and the increasing force-freeness of the field as a function of height. Of course the Lorentz force cannot be measured throughout the atmosphere, only at heights where magnetically sensitive spectral lines are formed. Practically, this restricts observational estimates of Lorentz forces to the thin layers of the photosphere up to the chromosphere. The height-dependence of the solar magnetic field is difficult to determine observationally because, while observations of chromospheric and coronal Zeeman lines are becoming more common ({\it e.g.}, Lin {\it et al.},~2004; Tomczyk {\it et al.},~2008; Rachmeler {\it et al.},~2013), the height of formation of the chromospheric lines can be ambiguous, and projection effects pose problems of interpretation for optically thin corona plasmas. Furthermore, detailed interpretation of asymmetric Stokes profiles is complicated by magnetic gradients and flows. Direct measurement of Zeeman splitting can provide tests of inversion methods but only for very strong fields.

The formation height of the Na~D line is believed to be about 600~km, a few pressure scale heights above the formation height of the photospheric Fe~I lines. From Mees Solar Observatory Na~D data Metcalf {\it et al.}~(1995) estimated a magnetic field gradient of about 0.8-1.1~G~km$^{-1}$ (1~G=1~gauss), which seems to be a typical estimate for the first Mm above the photosphere (Gary,~2001). From prominence magnetic field observations, Harvey~(1969) found an exponential field gradient of about 1.6~G~km$^{-1}$, representing a height range of about 10~Mm, above which height the field is believed to decrease in strength more steeply~(Gary,~2001).

Guided by these observations, Gary~(2001) represented the height profile of the sum of parametric dipole terms (Aschwanden {\it et al.},~1999),

\begin{equation}
B = B_{\rm s} \left/ \left( 1+\frac{z}{H_{\rm s}}\right)^3\right. + 
    B_{\rm f} \left/ \left( 1+\frac{z}{H_{\rm f}}\right)^3\right. + 
    B_{\rm w} \left/ \left( 1+\frac{z}{H_{\rm w}}\right)^3\right. ,
\label{eq:bprofile}
\end{equation}

\noindent where $H_{\rm s} = 0.5$~Mm, $H_{\rm f} = 75$~Mm, and $H_{\rm w} = 696$~Mm. The parameter $B_{\rm s}$ imposes a typical field strength in the photosphere, $B_{\rm f}$ represents the field strength in the mid-corona and $B_{\rm w}$ the global field strength taking into account the effect of the solar wind. The first term decays rapidly due to the expansion of the field in the chromospheric canopy. Gary~(2001) derived upper and lower envelopes for the field model representing sunspot and plage fields using the parameter values $B_{\rm s} = 2500$~G for the sunspot field and 100~G for the plage field, $B_{\rm f} = 50$~G for both fields, and $B_{\rm w} = 1$~G for the sunspot field and 0.005~G for the plage field. Figure~\ref{fig:gary} shows plots of four models described by Equation~(\ref{eq:bprofile}): the sunspot model (solid curve) the plage model (dot-dash curve) and two intermediate models with photospheric field strengths $B_{\rm s} = 1$~kG (dashed curve) and 500~G (dotted curve). We have set $B_{\rm f} = 50$~G and $B_{\rm w} = 1$~G in all four of our models since these constants represent the mid-coronal and global-coronal field strengths above the same active region in all four cases. We only need to vary $B_{\rm s}$ to compare the stratification of the sunspot umbral, plage and intermediate fields in our model region.

Equation~(\ref{eq:bprofile}) represents only dipole-like behavior of active region fields whereas active regions often have complex, multipolar photospheric flux distributions. Assuming that a region has approximately balanced magnetic flux and therefore has negligible monopole component, the dipolar component of this flux is the component that decreases most slowly with height and has the largest global influence. The dipole model therefore represents the field of most interest in our study, that most likely to contribute to the lateral and top boundary integrals in a Lorentz force calculation based on Equation~(\ref{eq:gaussdiv}).

Figure~\ref{fig:gary} (top) shows how the magnetic field strength $B=|{\bf B}|$ varies as a function of height for each of the four magnetic field models. For all four models the magnetic field strength decreases steeply in the first several Mm, showing the dominant influences of the $B_{\rm s}$ term in the first Mm and of the $B_{\rm f}$ term from there up to about 75~Mm. We will use these models to estimate the height-dependence of the Lorentz force according to Equation~(\ref{eq:gaussdiv}).

Gary's~(2001) plasma pressure model is formed by a linear combination of barometric terms representing chromospheric and coronal temperature regions,

\begin{equation}
p(z) = p_{\rm c} e^{-(z/H_{\rm c})(R/R_{\rm s})} + p_{\rm k} e^{-(z/H_{\rm k})(R/R_{\rm s})} ,
\label{eq:pprofile}
\end{equation}

\noindent where $R=R_{\rm s}+z$ is the radial coordinate, the sum of the solar radius $R_{\rm s}$ and the height $z$ above the solar surface, $H_{\rm k}$ is the chromospheric scale height, and $H_{\rm c} = H_0 (R/R_{\rm s})^2$ is the spatially varying effective coronal scale height corrected for gravity. The parameter values used are $p_{\rm c}=1.5$~dyn~cm$^{-1}$, $p_{\rm k}=1\times 10^5$~dyn~cm$^{-1}$, $H_0=55$~Mm and $H_{\rm k}=0.12$~Mm. This pressure profile was constrained by pressure estimates from Gary \& Alexander~(1999) derived from {\it Yohkoh}/SXT observations above an active region at the limb, and electron number density data corresponding to height $2.5\times 10^3$~Mm at temperature $T=2$~MK from Allen~(1973).

Figure~\ref{fig:gary} (bottom) shows how the plasma $\beta$, the ratio of the plasma and magnetic pressures given by Equation~(\ref{eq:beta}), varies as a function of height for each of the four magnetic field models. The plasma $\beta$ is around 1 (0.8) at the photosphere for the umbral field model (2500~G) and $> 1$ for the other models. The plage field model has plasma $\beta > 200$ at $z=0$~Mm. Within 1~Mm all four models have $\beta < 1$. At a height of about 100~Mm all models transition to a $\beta > 1$ regime. This means that there is a layer of the atmosphere in all four models between heights 1~Mm and 100~Mm where the field is dominant over the plasma and, having no significant plasma forces opposing it, is likely to be approximately force-free.

\section{Estimating the Lorentz Force Integral Contributions from the Different Atmospheric Layers}
\label{sect:lfintegral}

\begin{figure} 
\begin{center}
\resizebox{\textwidth}{!}{\includegraphics*{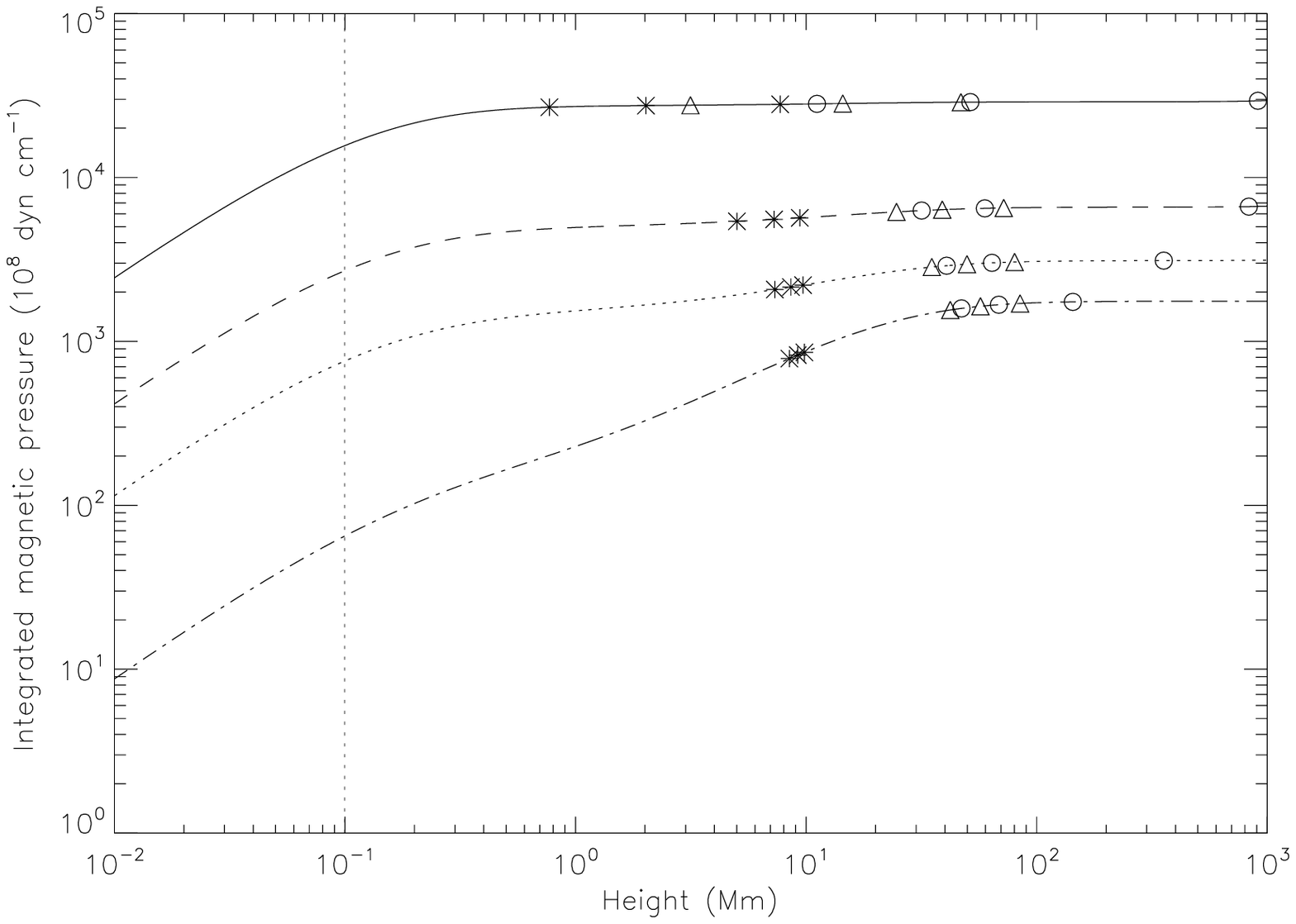}}
\resizebox{\textwidth}{!}{\includegraphics*{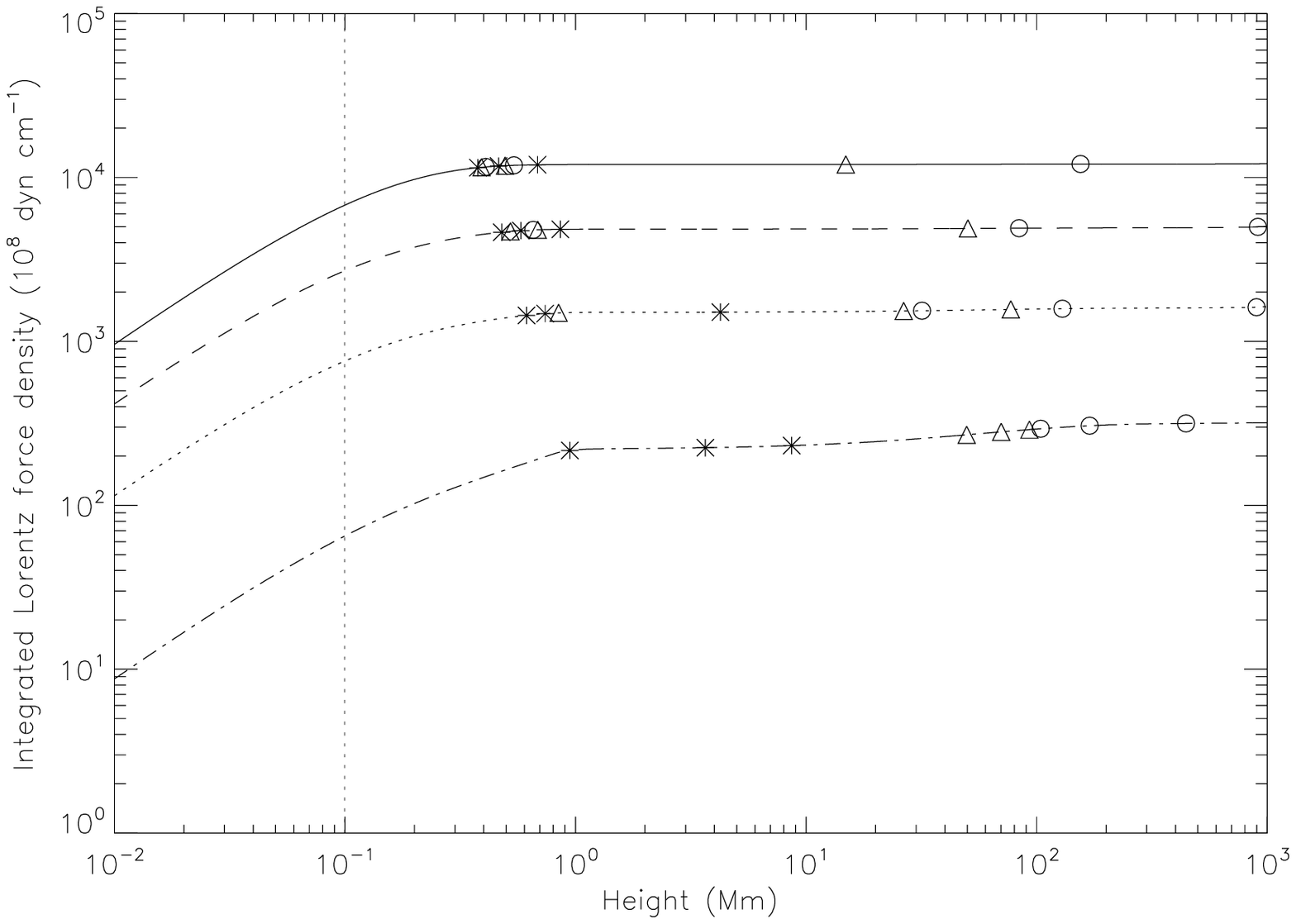}}
\end{center}
\caption{The integrated magnetic pressure (top) and integrated Lorentz force density (estimated Maxwell stress, bottom) for the four magnetic field models, plotted against height in the solar atmosphere. These quantities are integrated from $z=0$ to the heights indicated by the abscissa (see Equations~(\ref{eq:magpressure}) and (\ref{eq:lfestimate})), giving profiles that increase with height and become flat at great heights. For each model, the Lorentz force is estimated using the plasma pressure and $\beta$ as in Equation~(\ref{eq:lfestimate}). The solid and dot-dashed lines represent the models for sunspot and plage fields with photospheric field strengths of 2500~G and 100~G presented by Gary~(2001). The dashed and dotted lines represent intermediate fields of photospheric field strength 1000~G and 500~G.}
\label{fig:lfintegral}
\end{figure}

\begin{figure} 
\begin{center}
\resizebox{\textwidth}{!}{\includegraphics*{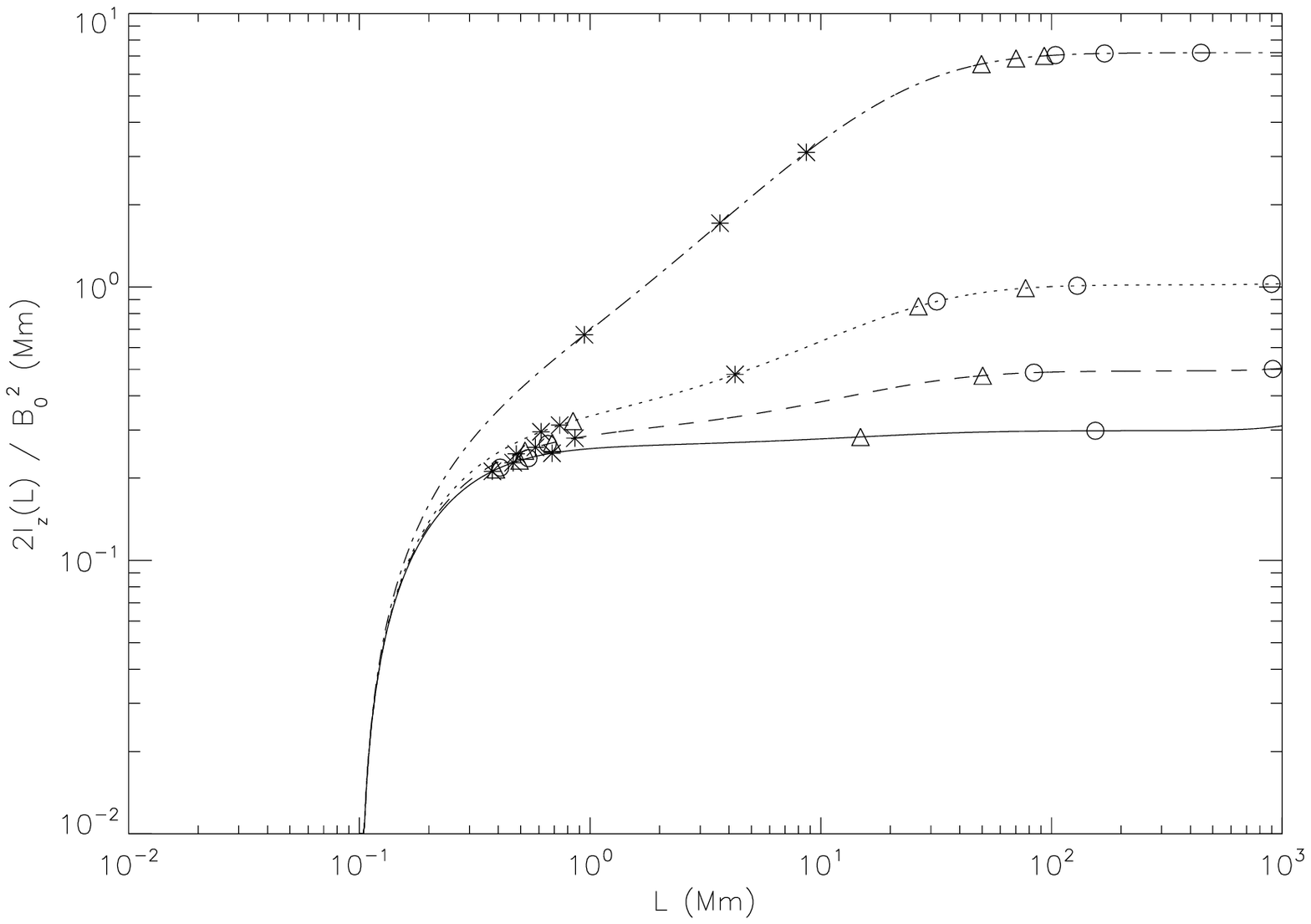}}
\resizebox{\textwidth}{!}{\includegraphics*{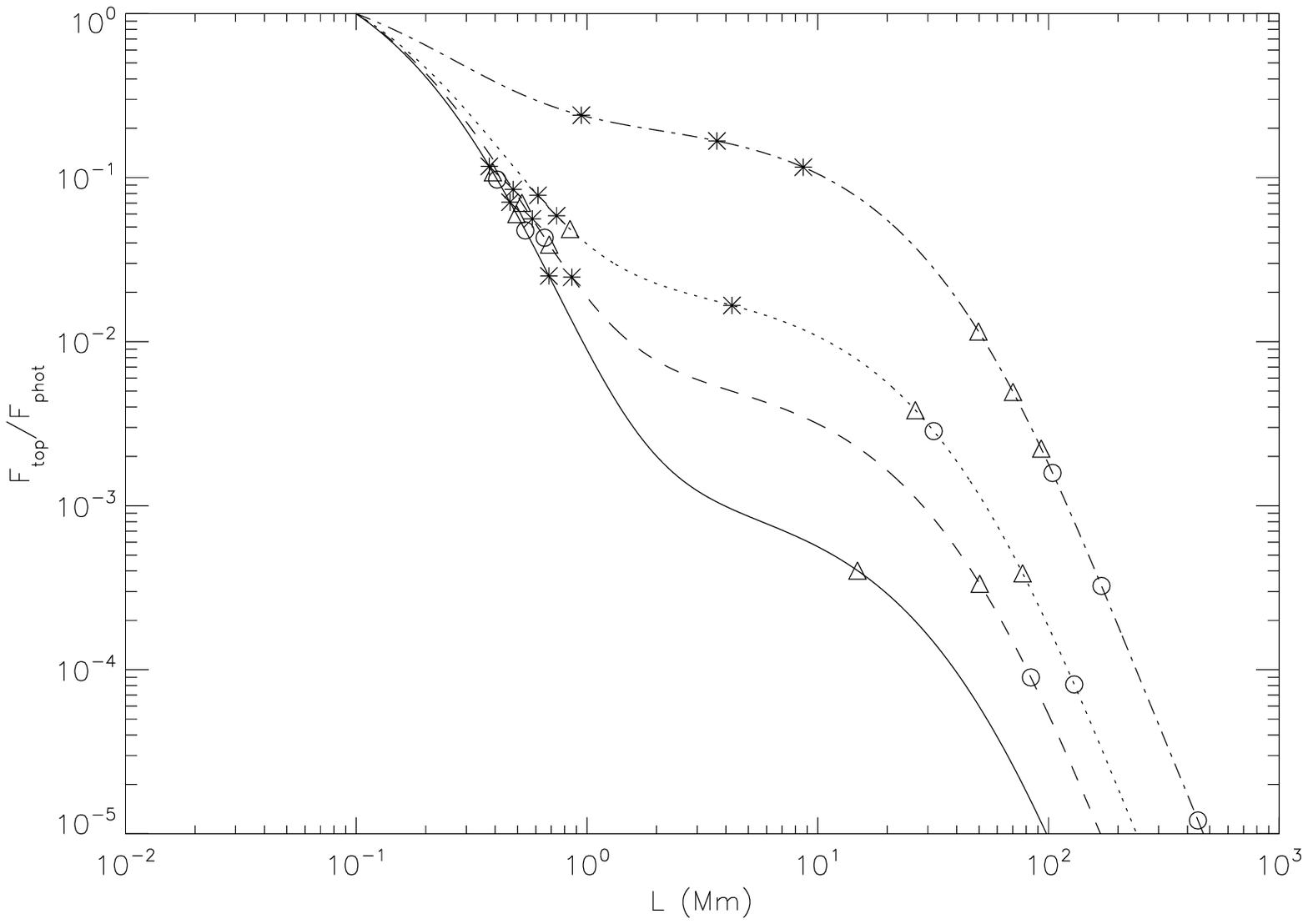}}
\end{center}
\caption{Comparison of the contributions from the photospheric boundary and the lateral boundaries (top) and comparison of the contributions from the photospheric boundary and the top boundary (bottom), plotted against top boundary heights $L$. The solid and dot-dashed lines represent the models for sunspot and plage fields with photospheric field strengths of 2500~G and 100~G presented by Gary~(2001). The dashed and dotted lines represent intermediate fields of photospheric field strength 1000~G and 500~G. In the top plot, for a given model and top boundary heights $L$, the field's horizontal length scale $L_B$ would have to be significantly larger than the value plotted for the lateral boundary contributions to be negligible. In the bottom plot the top boundary contribution is negligible wherever the curve is much less than one.}
\label{fig:bdycomp}
\end{figure}

The integral $P_{\rm m}$ of the magnetic pressure over height,

\begin{equation}
P_{\rm m}=\int_{0}^{z} B^2/8\pi\ {\mathrm d}z^{\prime}, 
\label{eq:magpressure}
\end{equation}

\noindent gives a crude estimate of the Lorentz force per unit area of the photosphere, approximating the left equality of Equation~(\ref{eq:gaussdiv}). This integral represents the magnetic pressure only. We ignore for the moment the approximately force-freeness of much of the coronal field, in which the opposition of the magnetic tension force to the magnetic pressure gradient force plays a crucial role. Figure~\ref{fig:lfintegral} (top) shows how this integral varies as a function of height for each of the four magnetic field models. In each case $P_{\rm m}$ increases monotonically to an asymptotic value at great heights but the rate of increase is different for the four field models. Whereas for the umbral field model $P_{\rm m}$ integrated up to 1~Mm almost matches its integral up to 1000~Mm, for the plage field model there is a an order of magnitude difference between the $P_{\rm m}$ integrals for the two heights. This suggests that the contribution to the magnetic pressure integral $P_{\rm m}$ integral above around 1~Mm may be negligible in the case of umbral fields but not for plage fields. For the intermediate field models there are significant contributions to the $P_{\rm m}$ integral up to about 10~Mm.

Here $z=0$ represents the height where the opacity $\tau =1$ at 5000~\AA , the usual definition of the photospheric surface. The vertical dotted lines in Figure~\ref{fig:lfintegral} indicate the approximate height $z=100$~km in the photosphere from where typical HMI observations derive according to Fleck, Couvidat, and Strauss~(2011). From correlations between simulated Doppler velocities and derived Dopplergrams, Fleck, Couvidat, and Strauss~(2011) concluded that the HMI observations of the Fe~{\sc i} 6173~\AA\ line originate from heights lower than the line core formation height, about 100~km. They found a similar result for the MDI signal, around 125~km, for the Ni~{\sc i} line at 6768~\AA\ also used by GONG. One may expect a similar result for the Fe~{\sc i} line at 6302.5~\AA\ used by {\it Hinode} and the NSO {\it Synoptic Optical Long-term Investigations of the Sun} (SOLIS) Vector Spectro-magnetograph (VSM), though the calculation has not been performed for this line. Fleck, Couvidat, and Strauss~(2011) also estimated that smearing effects associated with the instrumentation make the fine structure in the images appear to correspond to heights about 40-50~km higher in the photosphere. However, for the large, well resolved magnetic structures that we are considering, sunspot and neutral-line structures much larger than the HMI pixel size $0.\!\!^{\prime\prime}5$, the smearing effect is unlikely to be important. We therefore adopt the height $h_0=100$~km as the photospheric boundary for our Lorentz force estimates. We repeated the following analysis with the value $h_0=125$~km and found similar results.

Over-plotted on each curve in Figure~\ref{fig:lfintegral} (top) are three asterisks, three triangles and three circles. The three asterisks on each curve represent the heights $L$ where

\begin{equation}
\frac{1}{8\pi}\int_{z=h_0}^{L} B^2 {\rm d} z = \frac{C}{8\pi}\int_{z=L}^{h_1} B^2 {\rm d} z ,\label{eq:b2csymb}
\end{equation}

\noindent for the three cases with $C=0.9$, 0.95, and 0.99, and $h_1=10$~Mm. This value of $h_1$ corresponds to the typical height of a flare hard X-ray (HXR) source found in a statistical study of {\it Yohkoh} data by Matsushita {\it et al.}~(1992). Thus, on each curve, the first asterisk indicates the height $z=L$ where 90\% of the total magnetic pressure between the height of the HMI observations ($z=h_0$) and the height of a typical HXR source ($z=h_1=10$~Mm) is confined to heights below $z=L$. The second and third asterisks show the heights $L$ that contain 95\% and 99\% of the magnetic pressure. The three triangles show the equivalent heights $L$ for $h_1=100$~Mm, a typical heigh of a tall active region loop. The three circles show the equivalent heights $L$ for $h_1=1000$~Mm, which roughly represents the height high in the corona where the expanding solar wind dominate the magnetic field. The plot shows that, while most of the magnetic pressure associated with each of the four models is concentrated low in the atmosphere, $L$ needs to be at least several Mm in size for any but the strongest field models to concentrate more than 90\% of the magnetic pressure below $z=L$.

The magnetic pressure surely overestimates the size of the Maxwell stress and Lorentz force in the corridor between approximately $z=1$~Mm and $100$~Mm where the plasma $\beta < 1$ and the field is likely to be approximately force-free, with the magnetic pressure gradient force opposed by the magnetic tension force. In a nearly force-free field the Maxwell stress is not well approximated by the magnetic pressure alone. To estimate the effect of this force-freeness on the Lorentz force integral we modify the integral Equation~(\ref{eq:magpressure}) to become,

\begin{eqnarray}
L_{\rm f}=\int_{0}^{z} \beta_{<1} B^2/8\pi\ {\mathrm d}z^{\prime},
\label{eq:lfestimate}
\end{eqnarray}

\noindent using the factor

\[ \beta_{<1} =\left\{ \begin{array}{ll}
      \beta & \mbox{wherever $\beta < 1$} \\
      1 & \mbox{wherever $\beta \ge 1$}
                                  \end{array}
\right. \]

\noindent to simulate the varying influence of force-freeness as a function of height. This forces the integrand of Equation~(\ref{eq:lfestimate}) to equal the magnetic pressure $B^2/8\pi$ where $\beta < 1$ and the plasma pressure $p$ where $\beta > 1$. The reason for defining Equation~(\ref{eq:lfestimate}) in this way is as follows. If $\beta > 1$ then the magnetic field is dominated by the plasma so that the Maxwell stress has the typical size $B^2/8\pi$ of the magnetic pressure. If $\beta < 1$ then the magnetic field dominates the plasma, and much of the Maxwell stress is contained within the magnetic field: the magnetic pressure gradient can be opposed not only by the plasma pressure gradient but also by the magnetic tension force. The resulting Lorentz force is then smaller than the magnetic pressure gradient force. Since the plasma pressure gradient force is expected to balance this Lorentz force we set these forces to have equal size wherever $\beta < 1$. In cases with $\beta\ll 1$ the magnetic pressure and tension forces almost cancel and the field is approximmately force-free. We model this relationship between low plasma-$\beta$ values and the force-freeness of the field using the factor $\beta_{<1}$ in Equation~(\ref{eq:lfestimate}).

This simple model for the Lorentz force does not capture the complexity of force vectors in active regions. It is designed to describe the general size of the Lorentz force at different heights. In reality the direction of this vector can become an important consideration. For example, if a component of the Lorentz force vector becomes small at some location in the photosphere then contributions to Equation~(\ref{eq:gaussdiv}) from higher in the atmosphere might become significant at this location. However, such contributions would be small compared to the contributions from the photosphere generally and would not obscure the overall photospheric Lorentz force pattern. 

Figure~\ref{fig:lfintegral} (bottom) shows how this modified integral Equation~(\ref{eq:lfestimate}) for the integrated Lorentz force $L_{\rm f}$ varies as a function of height for each of the four magnetic field models, and comparing the two panels of Figure~\ref{fig:lfintegral} reveals the effect of the varying plasma $\beta$. The  integrals have become slightly smaller and the curves between $z=1$~Mm and $100$~Mm, where the plasma $\beta < 1$, are significantly flatter for the $L_{\rm f}$ integrals (bottom panel) than for the $P_{\rm m}$ integrals (top panel). The contributions to the $L_{\rm f}$ integral Equation~(\ref{eq:lfestimate}) from the nearly force-free fields between these heights is small enough to be negligible for the umbral model and the intermediate 1000~G model. For the 500~G model the integral up to 1~Mm is about 90\% of the integral up to 100~Mm.

Figure~\ref{fig:lfintegral} (bottom) also shows three asterisks, triangles and circles on each curve. These symbols correspond to heights $L$ where,

\begin{equation}
\frac{1}{8\pi}\int_{z=h_0}^{L} \beta_{<1} B^2 {\rm d} z = \frac{C}{8\pi}\int_{z=L}^{h_1} \beta_{<1} B^2 {\rm d} z ,\label{eq:b2betacsymb}
\end{equation}

\noindent where $h_1 = 10$~Mm (asterisks), 100~Mm (triangles), and 1000~Mm (circles), and the three symbols in each case represent $C=0.9$, 0.95, and 0.99 as in the top panel of Figure~\ref{fig:lfintegral} discussed above. It is clear from the bottom plot that, taking into account only fields up to 10~Mm (asterisks), over 90\% of the estimated Lorentz force is concentrated in $h_0\le z\le L$ with $L \approx 1$~Mm for all four models. If we relax our conditions to include all fields up to 100~M (triangles), then we still capture over 90\% of the estimated Lorentz forces within $h_0\le z\le L$ with $L < 1$~Mm for the three strongest models, excluding only the model for plage fields (dot-dashed curve). If we consider all fields up to 1000~Mm then the two strongest models still contain more than 90\% of their estimated Lorentz forces within the layer $h_0\le z\le L$ with $L \approx 1$~Mm. For $h_1=100$~Mm over 95\% of the two strongest models' estimated Lorentz forces are contained within $h_0\le z\le L$ with $L \approx 1$~Mm, and for $h_1=10$~Mm this percentage goes up to over 99\%. These results indicate that if we set $L\approx 1$~Mm and neglect all Lorentz force contributions above that height, then we can expect to miss less than 10\% associated with active region loops with $>500$~G photospheric fields ({\it i.e.}, $>500$~G at the $\tau =1$ opacity surface at 5000~\AA ), and less than 5\% for loops with $>1$~kG fields. A value $L\approx 1$~Mm therefore seems to be suggested by these considerations. Now we need to check the sizes of the contributions to Equation~(\ref{eq:gaussdiv}) from the different boundaries as discussed in Section~\ref{sect:subdomain}.

Figure~\ref{fig:bdycomp} allows a visual comparison between the contributions to the Lorentz force surface integral Equation~(\ref{eq:gaussdiv}) from the photospheric boundary and the lateral boundaries (top plot) and the top boundary (bottom plot). The top plot shows the quantity $2 I_z (L) / B_0^2$ for each of the four field models as a function of top boundary height $L$. For any value of $L$, the horizontal magnetic spatial scale $L_B$ of the structure must be comfortably larger than this quantity if the lateral boundary contribution $F_{\rm lat}$ is to be negligible - see Equation~(\ref{eq:latcomp}). The three asterisks, triangles and circles per curve in Figure~\ref{fig:bdycomp} (top) represent the values of $L$ shown in Figure~\ref{fig:lfintegral} (bottom) for the estimated Lorentz force. The cluster of symbols near and below $L=1$~Mm is concentrated between values $2 I_z (L) / B_0^2 = 200$~km and 300~km. The plot therefore shows that for the size of $L \approx 1$~Mm discussed above, $L_B$ must be significantly larger than 200~km or 300~km. Since this scale is significantly smaller than the typical length scales of the sunspots and major neutral lines generally studied using this method ({\it e.g.}, Wang {\it et al.},~2012; Petrie, 2012, 2013), this result indicates that the contributions to Equation~(\ref{eq:gaussdiv}) from the lateral boundaries, $F_{\rm lat}$, can be neglected for such structures, particularly those whose fields have strength corresponding to the two strongest models ($|{\bf B}| \ge 1$~kG at $z=0$).

Figure~\ref{fig:bdycomp} (bottom) shows the ratio between the top and photospheric  boundary contributions to Equation~(\ref{eq:gaussdiv}), $F_{\rm top} / F_{\rm phot}$, as a function of top boundary height $L$ for each model. For values of $L$ where this ratio is $\ll  1$, the contribution $F_{\rm top}$ to Equation~(\ref{eq:gaussdiv}) can be neglected. The three asterisks, triangles and circles per curve in this plot again represent the values of $L$ shown in Figure~\ref{fig:lfintegral} (bottom) for the estimated Lorentz force. For the cluster of symbols near and below $L=1$~Mm, the ratio $F_{\rm top} / F_{\rm phot}$ is between $10^{-2}$ and $10^{-1}$, indicating that the contribution $F_{\rm top}$ can also be neglected from estimates based on Equation~(\ref{eq:gaussdiv}) for our magnetic structures of interest.

There is some tension between Equations~(\ref{eq:latcomp}) and (\ref{eq:topcomp}) which can be seen by comparing the two plots of Figure~\ref{fig:bdycomp}. One of the ratios plotted in the figure cannot decrease without the other increasing: if the upper boundary height $L$ increases, the top boundary integral decreases but the lateral boundary integral must increase. On the other hand, stratification supports both inequalities in the same way, concentrating the forces near the lower boundary and making the lower boundary integral more dominant. The most stratified models (solid and dashed lines) therefore have the smallest ratios in both panels of Figure~\ref{fig:bdycomp}.

In Figure~\ref{fig:bdycomp} we have again represented the Maxwell stress tensor Equation~(\ref{eq:maxwelltensor}) by the magnetic pressure. The symbols of interest in Figure~\ref{fig:bdycomp} appear at heights below $z=1$~Mm and mostly correspond to plasmas with $\beta\ > 1$, where the magnetic pressure gradient likely represents the size of the Lorentz force reasonably well. In the strongest models $\beta$ becomes $< 1$ at heights $z < 1$~Mm. In these cases the Lorentz force likely becomes smaller than the magnetic pressure gradient where $\beta < 1$ because the magnetic tension generally must help the plasma pressure gradient force to oppose the magnetic pressure gradient force. Since this effect becomes larger as the height increases, the fall-off of the Lorentz force with height is therefore likely to be underestimated for the strongest models shown in Figure~\ref{fig:bdycomp}. However, the equivalent plots of Figure~\ref{fig:bdycomp} with $B^2$ replaced by $\beta_{<1} B^2$ (not shown) produced little effect on the cluster of symbols near $L=1$~Mm, suggesting that our simple model based on the magnetic pressure $B^2/8\pi$ adequately represents the Maxwell stress tensor within the height range of interest.

\section{Conclusion}
\label{sect:conclusion}

\begin{figure} 
\begin{center}
\resizebox{\textwidth}{!}{\includegraphics*{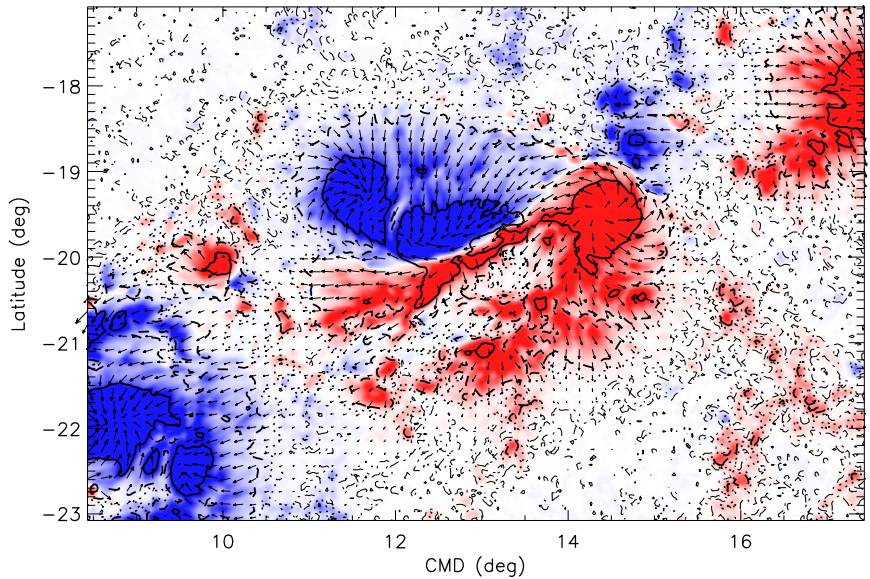}}
\end{center}
\caption{Vector magnetic field before the 15 February 2011 X2.2 flare. The vertical field component, $B_r$, is indicated by the color scale and the horizontal component by the arrows, with saturation values ±1000 G. Red/blue coloring represents positive/negative vertical field. The solid, dashed, dotted and dot-dashed contours indicate photospheric field strengths corresponding to the four models featuring in the previous figures and described in Section~\ref{sect:gary}. At the approximate height of the HMI observations, $z=100$~km, these models have field strength 1500~G, 629~G, 340~G, and 108~G, respectively.}
\label{fig:feb15flare}
\end{figure}

We conclude that for strong magnetic fields the contribution to the Lorentz force estimate from fields at height above around 1~Mm can be neglected. For photospheric subdomains filled with large, coherent structure and strong magnetic field, the contributions to the surface integrals derived by Fisher {\it et al.}~(2012) are negligible compared to the contribution from the photospheric boundary. This implies that Fisher {\it et al.}'s~(2012) method for estimating Lorentz force changes on the photosphere during solar flares using Equations~(\ref{eq:fr}) and (\ref{eq:fh}) can be applied to subdomains within a flaring active region, so long as the conditions on strength of field and coherence of structure are met within this photospheric subdomain. For HMI data, and for data originating from similar heights in the solar atmosphere, the horizontal length scale of the structure needs to be much larger than 300~km and the field needs to be stronger than about 630~G at the height of the observations ($\ge 1$~kG at the $\tau =1$ opacity layer at 5000~\AA). This conclusion therefore supports many of the published surface integral calculations based on Fisher {\it et al.}'s method, such as those by Petrie~(2012, 2013), who used regular rectangular integration domains covering strong neutral line fields and circular domains covering sunspots, and Wang and Liu~(2010) and Wang {\it et al.}~(2012), who used irregular domains covering strong sunspot and neutral line fields. Figure~\ref{fig:feb15flare}\footnote{This figure is updated from Figure~1 of Petrie~(2012). In Petrie~(2012) the longitude-latitude coordinate system has origin at the center of the remapped HMI magnetogram whereas here the origin corresponds to disk-center. See also http://hmi.stanford.edu/hminuggets/?p=539} shows the much-studied HMI vector magnetic field associated with the 15 February 2011 X2.2 flare. The plot shows that most of the important fields at the heart of the active region NOAA 11158 are strong enough to meet the above conditions, and the major features of the region have horizontal spatial scale much larger than 300~km.

Note that Fisher {\it et al.}'s~(2012) equations record only the Lorentz force difference between two vector magnetogram measurements covering the same locations at different times. During a highly dynamic event such as a flare, Lorentz forces apply to different layers of the solar atmosphere, but only those layers whose fields have significant and steady Lorentz force values before and after the flare will yield measurements that can be analyzed using this method. In a nearly force-free layer of the atmosphere the magnetic vectors may respond strongly to a flare. However, if the magnetic vector field in this layer is not accompanied by a significant and lasting Lorentz force change associated with the flare then Equations~(\ref{eq:deltafr}) and (\ref{eq:deltafh}) will not give a result that can be understood as a flare-related Lorentz force vector change. This is true even if the flare produced a large and permanent change in the magnetic vector field. In some nearly force-free layers of the solar atmosphere, dynamical equilibration processes remove any significant Lorentz forces from field configurations, erasing all traces of the Lorentz force changes produced by flares. It is therefore essential to the success of this method that the magnetogram observations derive from a layer where such force-removing equilibration processes do not generally occur.

\begin{acks}
I thank the referee for helpful comments. This project was prompted by discussions at Leverhulme Flare Seismology Workshop held at the Mullard Space Science Laboratory (MSSL), 9-12 September 2013. The author thanks the Leverhulme trust for hospitality during the author's visit to MSSL, and thanks his fellow participants at the workshop for interesting discussions during and after the meeting.
\end{acks}



 

\end{article} 

\end{document}